\documentclass[showpacs,preprintnumbers,amsmath,amssymb,prl,epsf]{revtex4}
\usepackage{graphicx}
\usepackage{dcolumn}
\usepackage{bm}

\newcommand{\be}{\begin{equation}}
\newcommand{\ee}{\end{equation}}
\newcommand{\bea}{\vspace{0.25cm}\begin{eqnarray}}
\newcommand{\eea}{\end{eqnarray}}

\def\PLA{{Phys. Lett.}  A }

\def\PRL{{Phys. Rev. Lett.} }

\begin{document}
\title{Experimental test of nonclassicality for a single particle}
\author{Giorgio Brida$^{1}$, Ivo Pietro Degiovanni$^{1}$,  Marco Genovese$^{1}$,
Valentina Schettini$^{1}$, \\ Sergey Polyakov$^{2}$ and Alan
Migdall$^{2}$}

\affiliation{$^{1}$Istituto Nazionale di Ricerca Metrologica, Strada
delle Cacce 91, 10135 Torino, Italy}

\affiliation{$^{2}$Optical Technology Division, National Institute
of Standards and Technology, 100 Bureau Drive, Gaithersburg, MD
20899-8441 and Joint Quantum Institute, Univ. of Maryland, College
Park, MD 20742}

\begin{abstract}

\begin{center}\parbox{14.5cm}
{In a recent paper [R. Alicki and N. Van Ryn, J. Phys. A: Math.
Theor., \textbf{41}, 062001 (2008)] a test of nonclassicality for a
single qubit was proposed. Here, we discuss the class of local
realistic theories to which this test applies and present an
experimental realization. }
\end{center}
\end{abstract}
 \maketitle \narrowtext

The quest for a classical theory able to reproduce the results of
Quantum Mechanics (QM) has a pluridecennial history, stemming from
the 1935 Einstein-Podolsky-Rosen paper \cite{Einstein}, where the
completeness of QM was questioned.

In 1964 Bell showed that for every local realistic theory (LRT)
\cite{Bell}, correlations among certain observables measured on
entangled states must satisfy a set of inequalities (the Bell's
inequalities, BI) while for QM they can be violated. In practice,
even though many experiments \cite{exp,prep} have shown violation of
BI's, their interpretation always calls for additional hypothesis
due to experimental limitations. In particular most of those results
rely on the accessory assumption that the events observed, due to
the finite efficiency of real detection apparatuses, are a faithful
statistical sample of the whole ensemble (this is often called {\em
detection loophole or fair sampling assumption}) \cite{prep,
Santos}. We note that while the detection loophole has been closed
for a ion system, in that case it has not been closed simultaneously
for space-like separated particles \cite{Monroe}. Hence, no
definitive experimental test of local realism has yet been performed
where all loopholes are closed simultaneously. In the last decade
the study of QM vs. LRT has attracted much interest fueled by the
development of Quantum Information science \cite{prep}. Furthermore,
the tests of certain realistic models are receiving much current
attention \cite{rm,prep}. In particular, some classes of LRTs have
not been excluded by Bell's inequality experiments because of
experimentally induced loophole(s). Experiments specifically aimed
at testing these LRTs are the focus of recent interest. Other
differences between quantum and classical treatments have also been
discovered and pointed out \cite{aul}.

Recently, a test of nonclassicality at single qubit level was
proposed \cite{al}.  This test is very appealing both because of its
simplicity (particularly in comparison with other proposals to test
nonclassicality at a single qubit level \cite{sq}) and its ability
to differentiate between a classical and a quantum state in a two
dimensional Hilbert space. Unfortunately, this procedure does not
test against all conceivable LRTs, and thus is not a general test of
nonclassicality. It does, however, allow testing of specific
classical models (i.e. these satisfying certain ``classical"
properties discussed later) against QM, although like the Bell test,
it is subject to the detection loophole depending on its
experimental implementation.

The purpose of this work is twofold: first, we want to start a
discussion on the advantages and limitations of this new proposal,
and second, we present the first experimental implementation of this
test, which we have realized with a conditional single-photon
source.

The proposal, \cite{al}, is based on the fact that given any two
positive real functions $\mathcal{A},\mathcal{B}$ obeying the
relation \be 0\leq\mathcal{A}(x) \leq \mathcal{B}(x) \label{HVT1}
\ee that for any probability distribution $\rho (x)$ it must be true
that \be \langle \mathcal{A}^2 \rangle \equiv \int \mathcal{A}^2(x)
\rho(x) dx \leq \int \mathcal{B}^2(x) \rho(x) dx \equiv
 \langle \mathcal{B}^2 \rangle. \label{HVT2}\ee

For quantum systems, one can find pairs of observables
$\widehat{A},\widehat{B}$ such that the minimum eigenvalue of
$\widehat{B} - \widehat{A}$ is greater than zero which we refer to
as the the inequality \be 0 \leq \widehat{A} \leq
 \widehat{B}. \label{classic} \ee
The commutation relations stemming from the classical approach that
lead to Eq. (\ref{HVT2}), prescribe that for all systems (described
by the density matrix $\widehat{\rho}$)  \be \langle \widehat{A}^2
\rangle \leq \langle \widehat{B}^2 \rangle, \label{qubitbell} \ee
where $\langle \widehat{O}
\rangle\equiv\mathrm{Tr}[\widehat{O}~\widehat{\rho} ]$, while to the
contrary, quantum theory allows that for certain quantum states \be
\langle \widehat{A}^2 \rangle
> \langle \widehat{B}^2 \rangle . \label{qubitbellViol}\ee
This sharp difference between classical (in the sense discussed
above) and quantum theory predictions at a single qubit level
\cite{footnote} can be tested experimentally on an ensemble of
single particles. In this paper we experimentally apply this method
to single-photons using the polarization degree of freedom.

We do note that because, by definition, hidden variables (such as
may be represented by $x$ above) cannot be observed directly, the
condition given in Eq. (\ref{HVT1}) defines and limits the class of
hidden variable theories that can be tested by violation of the
``classical" inequality (\ref{HVT2}) \cite{footnote1}.

Quantum objects used to implement this test are horizontally
polarized single-photons ($|H \rangle$) produced by a heralded
single-photon source. Our two observables are
\begin{equation}
\widehat{A}= a_{0}\widehat{P}_{\alpha } \end{equation} and
\begin{equation} \widehat{B}= b_{0}\left[p_{1}
\widehat{P}_{\beta}+ (1-p_{1})\widehat{P}_{\beta+\pi/2} \right],
\end{equation}
with numerical constants $a_{0}=0.74$ and  $b_{0}=1.2987$,
$\widehat{P}_{\theta }$ is the projector on the state
$|s(\theta)\rangle =\cos\theta |H\rangle +\sin\theta |V\rangle$ (and
$\widehat{P}_{\theta +\pi/2 }$ is the projector on the orthogonal
state $\sin\theta |H\rangle -\cos \theta |V\rangle$), and $0\leq
p_{1} \leq 1$.

The expectation value $\langle\widehat{A}\rangle$ can be obtained
experimentally by projecting heralded photons onto the state
$|s(\alpha)\rangle$, while $\langle\widehat{B}\rangle$ is realized
with an experimental setup that projects heralded photons onto the
state $|s(\beta)\rangle$ with probability $p_{1}$, and onto the
state $|s(\beta+\pi/2)\rangle$ with probability $(1-p_{1})$. This
probabilistic projection can be achieved, in principle, with a
beam-splitter with a splitting ratio $p_{1}$, sending photons
towards the two projection systems.

The experimental measurement of both $\langle\widehat{A}\rangle$ and
$\langle\widehat{A}^{2}\rangle$, where $\widehat{A}^{2}=a_{0}^{2}
\widehat{P}_{\alpha }$, is achieved by projecting the photon onto
the state $|s(\alpha )\rangle$. To measure
$\langle\widehat{B}^{2}\rangle$, where
$\widehat{B}^{2}=b_{0}^{2}\left[p_{1}^{2} \widehat{P}_{\beta}+
(1-p_{1})^{2}\widehat{P}_{\beta+\pi/2}  \right]$, however, it is
necessary to change the beam splitting ratio to
$p_{2}=\frac{p_{1}^{2}}{p_{1}^{2}+ (1-p_{1})^{2}} $. Thus the
operator $\widehat{B}^{2}$ is:
\begin{equation}
\widehat{B}^{2}=b_{0}^{2}\frac{1-2\sqrt{(1-p_{2})p_{2}}}{(1- 2
p_2)^2} \left[ p_{2} \widehat{P}_{\beta}+
(1-p_{2})\widehat{P}_{\beta+\pi/2} \right],
\end{equation}
in terms of the splitting ratio $p_{2}$. We assume that the
beamsplitter randomly and fairly splits the incoming photons, with
the probabilities that are equivalent to the splitting ratio of
``classical'' waves.

It can be shown that for the parameters set to $p_{1}=4/5$,
 $p_{2}=16/17$, $\alpha=11/36 ~ \pi $, and
$\beta=5/12 ~ \pi$, the results predicted by quantum theory are
$\langle \widehat{B}^{2} \rangle -\langle \widehat{A}^{2}
\rangle=-0.0449  $, and $\langle \widehat{B} \rangle -\langle
\widehat{A} \rangle= 0.0685$,  while the minimum eigenvalue of $
\widehat{B} - \widehat{A} $ is $d_{-} = 0.0189$, where
\begin{equation}\label{dminus}
d_{-}\equiv\frac{1}{2}\left\{b_0- a_0-\sqrt{a_0^2 +b_0^2~(1-2
p_1)^2+ 2~ a_0~ b_0~(1-2 p_1)~\cos[2(\alpha-\beta)] } \right\}.
\end{equation}

The critical question is whether the above arrangement can serve as
a test of all LRTs.
As mentioned in the discussion of Eqs. (1)-(5), the test proposed in
Ref. \cite{al} concerns the class of LRTs satisfying Eq. (1) only.
The simplest example of a LRT that does not satisfy this condition
and can mimic QM, relies on a hidden (or simply unmeasured) variable
$ x $ uniformly distributed between 0 and 1 ($\rho(x)=1$ when $0
\leq x \leq 1$), and classical quantities $\mathcal{A}(x)= a_{0}
\theta(X_{A}-x)$, and $\mathcal{B}(x)= b_{0} [p_{1}
\theta(X_{B}-x)+(1-p_{1})\theta(x-X_{B})]$, where $\theta(\xi)$ is
the step function (1 for $\xi\geq 0$, and 0 elsewhere). By choosing
$X_{A}= \cos^{2}\alpha$ and $X_{B}= \cos^{2}\beta$ and using the
experimental parameters defined above, we obtain the quantum
mechanical predictions,  $\langle \mathcal{B} \rangle -\langle
\mathcal{A} \rangle= 0.0685$ and $\langle \mathcal{B}^2 \rangle
-\langle \mathcal{A}^2 \rangle= -0.0449$. It is easy to verify that
this model does not belong to the class of hidden variable models
falsified by this test, as the condition given in Eq. (\ref{HVT1})
is not satisfied for some $ x $. In particular, for $X_{B}<x<X_{A}$
we have $\mathcal{A}(x) > \mathcal{B}(x)$. The boundary of the class
of LRTs identified by condition (\ref{HVT1}), as well as the
possibility of enlarging the class by modifying this method is a
very important question, but is beyond the scope of this work.

\begin{figure}[tbp]
\begin{center}
\includegraphics[angle=0,width=9 cm]{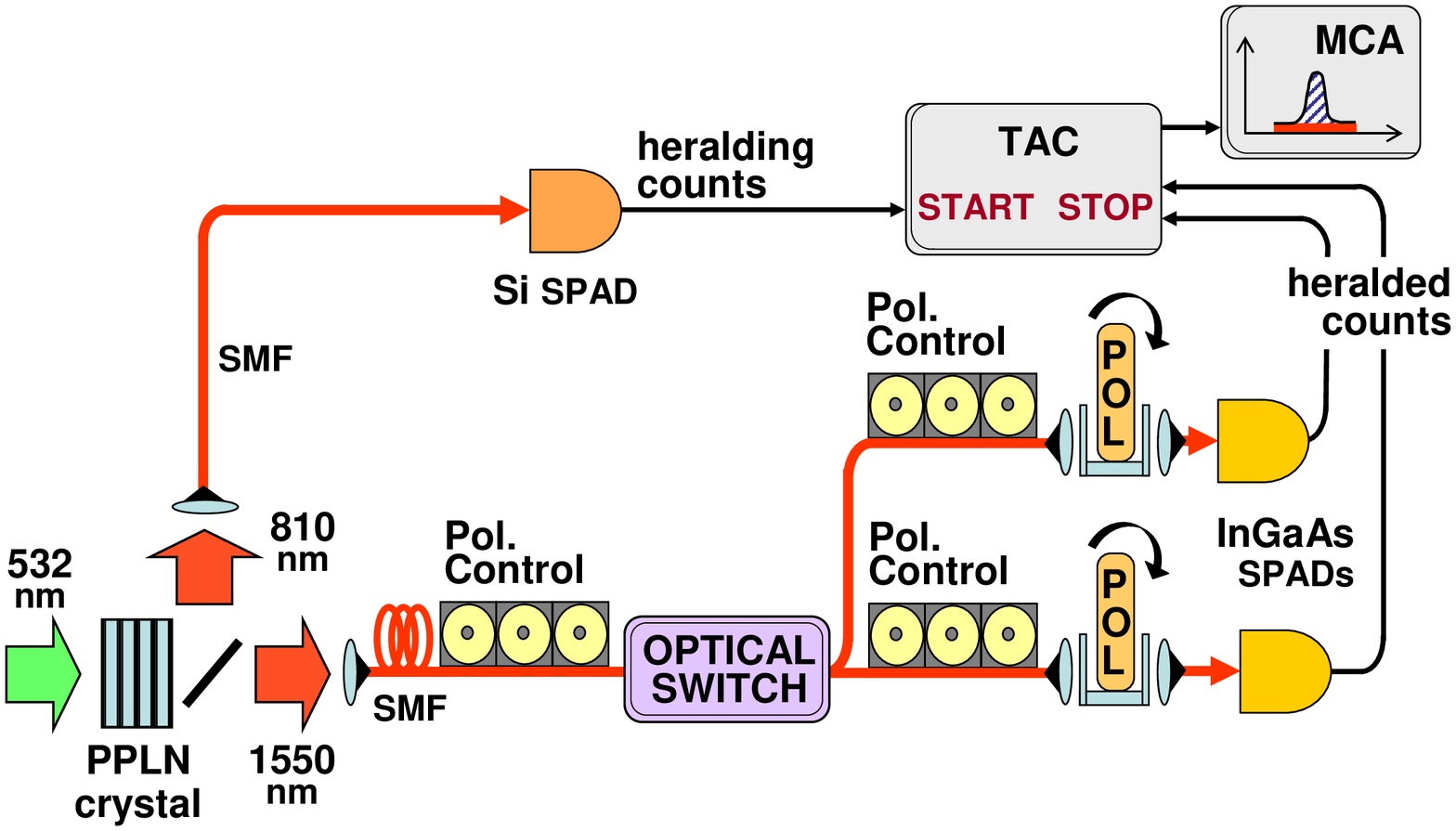}
\end{center}
\caption{Experimental setup. A PDC heralded single-photon source
generates pairs of photons at 810 nm (heralding) and 1550 nm
(heralded) in a PPLN crystal pumped by a 532 nm laser. The heralded
photons are sent to the measurement apparatus designed to evaluate
the observables $\langle\widehat{A}\rangle$,
$\langle\widehat{A}^{2}\rangle$, $\langle\widehat{B}\rangle$, and
$\langle\widehat{B}^{2}\rangle$.} \label{setup}
\end{figure}

The experimental setup is presented in Fig. \ref{setup}. The
heralded single-photon source is based on photon pairs produced by
parametric down conversion (PDC).  Our PDC source is a 5 mm long
periodically poled MgO-doped lithium niobate (PPLN) crystal, pumped
by a continuous wave (cw) laser at 532 nm, that produces pairs of
correlated photons at 810 nm and 1550 nm \cite{metrologia}. A cutoff
filter blocks the pump laser light at the crystal's output and a
dichroic mirror separates the 810 nm and 1550 nm photons. Extra
interference filters at 810 and 1550 nm with a full width
half-maximum (FWHM) of 10 nm and 30 nm, respectively, further
suppress fluorescence from the PPLN crystal reducing background
counts. The collection geometry on the heralding arm restricts the
visible bandwidth to $\approx$2 nm FWHM.

The heralded single-photon source is independently characterized to
ensure that a) there is a sizable correlation between the signal
photons at 810 and 1550 nm that dominates over the background of
accidental coincidences and b) that multiphoton emission is
negligible (i.e. when conditioned on a photon detection at 810 nm,
the probability to observe two photons in a 1550 path is negligible,
see Appendix.)

For the experiment, photons of the heralding arm are routed by a
single-mode fiber (SMF) to a Si-single-photon Avalanche Diode (SPAD)
operating in Geiger mode, while photons in the heralded arm, coupled
into a second SMF, are sent to the apparatus that implements the
probabilistic projections according to the parameter values
determined above (necessary to measure $\langle \widehat{B} \rangle$
and $\langle \widehat{B}^{2} \rangle$). These projections are
implemented by means of an all-fiber variable beam splitter and
polarizers.

The variable beam-splitter is made from an optical switch that can
route heralded photons with an adjustable splitting ratio into two
different optical paths \cite{IEEE}. The input polarization state in
each optical path after the beamsplitter is controlled using a three
paddle single-mode fiber polarization rotator followed by a rotating
polarizer (POL).

This scheme allows us to experimentally set the polarizers to
perform projections on $\widehat{P}_{\alpha}$,
$\widehat{P}_{\beta}$, and $\widehat{P}_{\beta+\pi/2}$, and set the
splitting probability of the beam splitter to $p_{1}$ or $p_{2}$ to
make necessary measurements of the observables. After passing
through the polarizers that performed the projections, the heralded
photons were finally sent to InGaAs-SPADs gated by the Si-SPAD
heralding counts.

We determine the true coincidence probability for each gate, rather
than using the raw measured counts to eliminate the contribution of
accidental coincidences, detector deadtimes, and drifts. The
probability for each measurement $i$ was evaluated according to
\begin{equation} \label{eta}
\eta_{i}(\theta,p)=\frac{N_{i}(\theta,p)}{M_{g,i}},
\end{equation}
where $N_{i}(\theta,p)$ is the number of coincidences sent with
probability $p$ towards the detection system with the polarizer
projecting photons onto the state $|s(\theta)\rangle$, and $M_{g,i}$
is the number of heralding gate counts. Thus, in each experimental
configuration, $\langle \widehat{P}(\theta)\rangle$ was estimated as
\begin{equation} \label{eta}
\mathcal{E}[\langle\widehat{P}(\theta)\rangle]=\frac{\sum_{i}
\eta_{i}(\theta,p)}{\sum_{i}[\eta_{i}(\theta,p)+\eta_{i}(\theta+\pi/2,p)]},
\end{equation}
while the probability of sending a photon towards a detection system
(whose nominal value is $p$) was estimated as
\begin{equation} \label{p}
\mathcal{E}[p]=\frac{\sum_{i} [
\eta_{i}(\theta,p)+\eta_{i}(\theta+\pi/2,p)]}{\sum_{i}\left[
\begin{array}{c}
\eta_{i}(\theta,p)+\eta_{i}(\theta,1-p) +\\
\eta_{i}(\theta+\pi/2,p)+\eta_{i}(\theta+\pi/2,1-p)
\end{array} \right]}.
\end{equation}

\begin{table}
\caption{\label{tab:table1} Measurement results with statistical and
total uncertainties and theoretical predictions.}
\begin{ruledtabular}
\begin{tabular}{ccc}
Quantity & Measurement\footnote{The total uncertainty (in
parentheses) accounts for both statistical and systematic effects.
}& QM theory\\
\hline $\mathcal{E}[\langle \widehat{B}\rangle  - \langle
\widehat{A}\rangle ]$ & 0.0581 $\pm$ 0.0049 ($\pm$ 0.0112) & 0.0685\\
$\mathcal{E}[\langle \widehat{B}^{2}\rangle - \langle
\widehat{A}^{2}\rangle  ]$ & -0.0403 $\pm$ 0.0043 ($\pm$ 0.0066) & -0.0449\\
  &   & ($>0$ LRT)\\
  &  &    \\
 $\mathcal{E}[p_{1}]$ &  0.80 $\pm$ 0.01 & 0.800 \\
$\mathcal{E}[p_{2}]$ &  0.94 $\pm$ 0.01 & 0.941 \\
\end{tabular}
\end{ruledtabular}
\end{table}

Using Eqs. (\ref{eta}) and (\ref{p}) we computed the experimental
values of $\langle \widehat{A} \rangle$, $\langle \widehat{A}^{2}
\rangle$, $\langle \widehat{B} \rangle$, and $\langle
\widehat{B}^{2} \rangle$ as seen in Table I. From the same
experimental results we obtained an indirect evaluation of the
minimum eigenvalue of $\widehat{B} -
 \widehat{A}$ as $(0.0101\pm 0.0065)$, showing
that we have met requirement (3) (we point out that the high
relative uncertainty of this evaluation is due to its indirect
determination). From the value of $\mathcal{E}[\langle
\widehat{B}^{2}\rangle - \langle \widehat{A}^{2}\rangle  ]$ we show
a violation of the classical limit (for appropriate LRTs) by more
than 6 standard deviations.

Table I presents both the statistical and total uncertainties.
Statistical uncertainties include those due to Poisson counting
statistics as well as those due to random misalignment of the
polarizers (we estimate an angular uncertainty of 2.5$^{\circ}$),
while the total uncertainties also include systematic effects such
as the uncertainty in setting the optical switch voltage bias used
to obtain the required splitting ratio. As an additional test, we
measured $\mathcal{E}[p_{1}]$ and $\mathcal{E}[p_{2}]$ and found
them consistent with the intended settings (see Table I).
Furthermore, we analyzed how the non-ideal (multi-photon) behavior
of our single-photon source might have affected the experimental
results, and we found its effects to be negligible, being more than
an order magnitude below the listed uncertainties.

In conclusion, we have investigated the theoretical proposal for
testing nonclassicality of a single-particle state \cite{al}. While
the utility of this test is open to question, as it does not apply
to every conceivable LRT like Bell's inequalities, but only the
class of LRTs satisfying Eq. (\ref{HVT1}), we have nonetheless
experimentally implemented it as proposed. Following the test's
protocol, our measurement results are seen to be incompatible with a
certain class of LRTs (as defined by Eq. (1)) while being well
predicted by QM. In particular, our results clearly falsify this LRT
class by 6 standard deviations. The precise identification of this
class and whether and if it maps to any physical system remains to
be determined. Also to be determined is whether it is possible to
extend or generalize this test to cover a larger class of LRTs. This
effort represents a first step in this direction of providing a
sharp difference between QM and LRTs at the single qubit level/two
dimensional Hilbert space and a physical implementation of that
test.

\section{Appendix}

A necessary requirement for a convincingly realizing the Alicki-Van
Ryn's proposal \cite{al} is a demonstration that our source in fact
produces single-photon states.

\begin{figure}[tbp]
\begin{center}
\includegraphics[angle=0,width=9 cm]{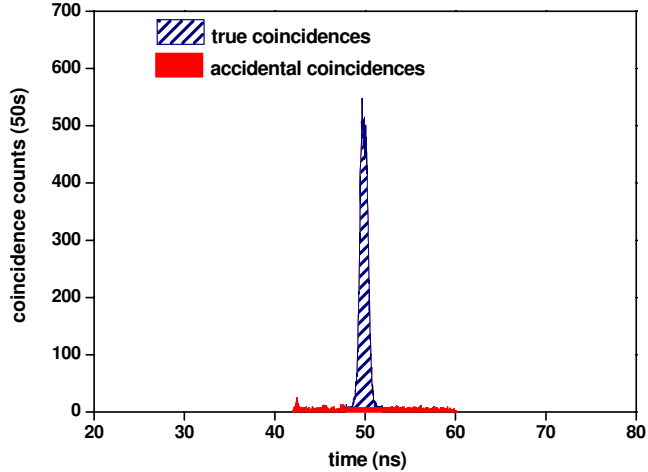}
\end{center}
\caption{Typical correlation between the detection of heralding and
heralded photons, showing the coincidences peak due to heralded
counts (true coincidences) and the uniformly distributed accidental
coincidences. InGaAs-SPAD gate time was approximately 20 ns.  }
\label{peak}
\end{figure}

First, we verify that the source optics are aligned to collect
correlated photons. The correlation between the two arms of the
source is measured with a  Time to Amplitude Converter (TAC) and a
Multi-Channel Analyzer (MCA). The MCA output (Fig. \ref{peak}) shows
the correlation peak along with the background of uniformly
distributed accidental counts, as expected for our photon source.
(We used a gate time of $\approx20$ ns for the InGaAs-SPAD.) From
this shape, we can subtract the background (i.e. counts not produced
by photons of the same pair) from signal, or true coincidences (i.e.
the simultaneous generation of a heralded photon and its heralding
twin).

Second, we verify that the possibility of having more than one
photon in the heralded arm after detecting the heralding photon is
low. With this aim we use the same setup as for the main experiment
(Fig. 1), but with the polarizers removed and the splitting factor
of the switch set to $p=(0.50 \pm 0.01)$. The efficiency of a
single-photon source can be described by means of the two parameters
$\Gamma_{1}= Q(\mathrm{1})/Q(\mathrm{0})$ and $\Gamma_{2}=
Q(\mathrm{2})/Q(\mathrm{1})$, where $Q(\mathrm{0})$ is the
probability that for each heralding count neither InGaAs-SPAD in the
heralded arm fires, $Q(\mathrm{1})$ is the probability of detecting
just one count for each herald, and $Q(\mathrm{2})$ is the
probability of observing a coincidence for each heralding count from
simultaneous firings by the two InGaAs-SPADs.

In general, a heralding detection announces the arrival of a
``pulse" containing $n$ photons at the heralded channel. The
probability of a specific InGaAs-SPAD firing due to a heralded
optical pulse containing $n$ photons is
\begin{eqnarray}\nonumber
Q(\mathrm{1}|n)&=&\sum_{m=0}^{n} [1-(1-\tau)^{m}]B(m|n;p)= \\
&=& 1-(1- p ~ \tau)^{n},
\end{eqnarray}
where $p$ is the optical switch splitting ratio, $B(m|n;p)=n![m!~
(n-m)!]^{-1}p^{m}(1-p)^{n-m}$ is the binomial distribution
representing the splitting of $n$ photons towards the two
InGaAs-SPADs, and $\tau$ is the detection efficiency of each
InGaAs-SPAD (that also accounts for all collection and optical
losses in the channel). Analogously, the probability of observing a
coincidence between the two InGaAs-SPADs due to a heralded optical
pulse with $n$ photons is
\begin{eqnarray}\nonumber
& Q(\mathrm{2}|n)= \\
&=\sum_{m=0}^{n} [1-(1-\tau)^{m}][1-(1-\tau)^{n-m}]B(m|n;p) \\
\nonumber &= 1-(1- p ~ \tau)^{n}-[1- (1-p)  \tau]^{n}+(1- \tau)^{n}.
\end{eqnarray}

Thus we get $Q(\mathrm{1})=\sum_{n} Q(\mathrm{1}|n)\mathcal{P}(n)$,
and $Q(\mathrm{2})=\sum_{n} Q(\mathrm{2}|n)\mathcal{P}(n)$ for
$\mathcal{P}(n)$ being the general probability distribution of the
number of photons in a heralded optical pulse. Setting $p=0.5$, in
the case of an ideal single-photon source
($\mathcal{P}(n)=\delta_{n,1}$) we obtain $Q(\mathrm{1})=\tau/2 $,
and $Q(\mathrm{2})=0$, corresponding to $\Gamma_{2}=0$ and
$\Gamma_{1}=\tau/[2 (1-\tau/2)]$; while for a Poissonian source
($\mathcal{P}(n)=\mu^{n} e^{-\mu}/n!$) we obtain
$Q(\mathrm{1})=1-\exp(-\tau \mu/2 )$, and
$Q(\mathrm{2})=[1-\exp(-\tau \mu/2 )]^{2}$, corresponding to
$\Gamma_{2}=1-\exp(-\tau \mu/2 )$ and $\Gamma_{1}=\exp(\tau \mu/2
)-1$ (meaning $\Gamma_{2}\simeq \Gamma_{1}=\tau \mu/2 $ when $\tau
\mu \ll 1$). See Table II for comparison between the ideal sources
above and our implementation.

\begin{table}
\caption{\label{tab:table1} Two-photon characterization of
single-photon source, Poisson source and our source without and with
background subtraction.}
\begin{ruledtabular}
\begin{tabular}{cccc}
 Source Type &  $\Gamma_{1}$ &
$\Gamma_{2}$ & $\Gamma_{2} / \Gamma_{1}$\\
\hline Single-photon & $\tau/ [2
(1-\tau/2)]$ & 0 & 0\\
Poisson   & $e^{\tau \mu/2 }-1$ & $1-e^{-\tau \mu/2 }$ & $\approx 1$
(when $\tau
\mu \ll 1$)\\
 This & $(4.14\pm 0.06)\cdot10^{-3}$ & $(0.66 \pm 0.06)\cdot10^{-3}$ & $0.16\pm 0.01$\\
  This (bkg subtr.) & $(4.02\pm 0.06)\cdot10^{-3}$ & $(0.37 \pm 0.36)\cdot10^{-3}$ & $0.09\pm 0.09$
\end{tabular}
\end{ruledtabular}
\end{table}

From our experimental data we obtained, with background subtraction,
results for $\Gamma_{2} $ that are compatible with 0 as for ideal
single-photon sources. We also note that $\mathcal{E}[\Gamma_{1}]$,
is in agreement with the estimated optical losses and a previous
detector calibration (Table 1) \cite{IEEE}.

An alternative characterization metric for single-photon sources,
was proposed by Grangier \textit{et al.} \cite{grangier}. They
introduced an ``anticorrelation criterion" based on the parameter
$\alpha=Q(\mathrm{2})/[Q^{(I)}(\mathrm{1}) ~ Q^{(II)}(\mathrm{1})]$
((I), (II) indicate the two detectors after the variable beam
splitter). For an ideal single-photon source $\alpha=0$, while
$\alpha \geq 1$ corresponds to classical sources. From our
experimental data $\mathcal{E}[\alpha]=(0.18 \pm 0.02)$ and
$\mathcal{E}[\alpha]=(0.11 \pm 0.11)$ with and without background
subtraction, respectively, ensuring that conditional single-photon
output dominates for our source.

\section{ Acknowledgments}
We thank R. Alicki, E. Knill and E. Tiesinga for helpful
discussions. This work has been supported in part by Regione
Piemonte (E14) and by the MURI Center for Photonic Quantum
Information Systems (Army Research Office (ARO)/ Intelligence
Advanced Research Projects Activity (IARPA) program
DAAD19-03-1-0199) and the IARPA entangled source programs.


\begin{references}
\bibitem{footnote1} A study to fully identify such a class and possibility of its expansion via additional measurements
is underway.
\end{references}
\end{document}